# Electrically powered locomotion of dual-nature colloid-hedgehog and colloid-umbilic topological and elastic dipoles in liquid crystals

*Bohdan Senyuk,[†] Richmond E. Adufu,[‡] and Ivan. I. Smalyukh\*[,†,‡,§,^,ǁ]*

[†]Department of Physics, University of Colorado, Boulder, Colorado 80309, United States

[‡]Department of Electrical, Computer and Energy Engineering, University of Colorado, Boulder, Colorado 80309, United States

[§]Soft Materials Research Center and Materials Science and Engineering Program, University of Colorado, Boulder, Colorado 80309, United States

[^]Chemical Physics Program, Departments of Chemistry and Physics, University of Colorado, Boulder, Colorado 80309, United States

[ǁ]Renewable and Sustainable Energy Institute, National Renewable Energy Laboratory and University of Colorado, Boulder, Colorado 80309, United States

**ABSTRACT:** Colloidal particles in liquid crystals tend to induce topological defects and distortions of the molecular alignment within the surrounding anisotropic host medium, which results in elasticity-mediated interactions not accessible to their counterparts within isotropic fluid hosts. Such particle-induced coronae of perturbed nematic order are highly responsive to external electric fields, even when the uniformly aligned host medium away from particles exhibits no response to fields below the realignment threshold. Here we harness the non-reciprocal nature of these facile electric responses to demonstrate colloidal locomotion. Oscillations of electric field prompt repetitive deformations of the corona of dipolar elastic distortions around the colloidal inclusions, which at appropriately designed electric driving synchronize the displacement directions. We observe the colloid-hedgehog dipole accompanied by an umbilical defect in the tilt directionality field (**c**-field), along with the texture of elastic distortions that evolves with changing applied voltage. The temporal out-of-equilibrium evolution of the director and **c**-field distortions around particles upon turning voltage on and off is not invariant upon reversal of time, prompting lateral translations and interactions that markedly differ from what is accessible to these colloids under equilibrium conditions. Our findings may lead to both technological and fundamental science applications of nematic colloids as both model reconfigurable colloidal systems and as mesostructured materials with pre-designed temporal evolution of structure and composition.

**INTRODUCTION**

Synthetic systems made of active or externally driven components have opened new perspectives for materials design and applications. Colloidal and liquid crystalline (LC) systems are examples of soft matter that often exhibit stark structural similarities with the forms of molecular and colloidal self-organization found in living organisms, but they are most commonly studied under equilibrium conditions.[1,2] Manipulation and motion of individual colloidal particles or microswimmers and their ensembles in different host media self-propelled or controlled by means of chemical, thermal, electrical or other external stimuli attract a great deal of attention due to the possible new ways of assembly, microfluidic transport, and variety of their collective dynamics reminiscent of schooling and flocking among living organisms, the active matter.[3-9] LCs as a host for such driven colloidal systems introduce another level of complexity due to the anisotropy of their physical properties with respect to the director field $\mathbf{n}(\mathbf{r}) \equiv -\mathbf{n}(\mathbf{r})$ describing the locally-averaged spatial patterns of orientation of LC molecules. This complexity comes along with variety of new possibilities and means of directing the motion of colloidal microswimmers with increased level of control due to facile response of LC molecules themselves to such external stimuli as magnetic, optical or electric fields.[10-12] Colloidal dynamics in LCs exhibit a rich diversity of physical phenomena[13] and, recently, a number of studies explored the electrically driven transport of colloidal particles[13-19] and particle-like topological solitons[8,9,20] in LCs.

Here we focus on electrically driven directional locomotion of artificial microswimmers formed by elastic colloidal dipoles[21-24] in nematic LCs. We show that electrically driven propulsion of elastic dipoles in a strongly confined LC results from the nonreciprocal director dynamics that can be also understood in terms of complex elastic interactions between colloidal particles and induced nonsingular umbilical defects that are reconfigured by voltage. We observe that upon application of electric field, the uniaxial colloid-hedgehog dipole[21,22,24] is accompanied by an additional umbilical defect in the $\mathbf{n}(\mathbf{r})$-tilt directionality field (the $\mathbf{c}$-field) forming a field-induced biaxial nematostatic dipole.[25,26] We describe how oscillations of specially designed low frequency electric field cause repetitive transformation of director distortions around the colloidal inclusion and this temporal out-of-equilibrium evolution of the LC director and **c**-field distortions around particles upon turning voltage on and off is not invariant upon reversal of time, prompting lateral locomotion of the colloidal particles. Our findings provide new insights into the phenomenological richness of colloids in LCs in terms of their controlled manipulation, assembly and collective dynamics.

**EXPERIMENTAL SECTION**

  **Materials.** We used a room temperature nematic LC ZLI-2806 (EMD Electronics) as a host medium for colloidal particles. To obtain elastic dipoles, we used silica spheres with a diameter $2R_0$=3.14 $\mu$m (Duke Scientific) treated with an aqueous solution (0.05 wt%) of N,N-dimethyl-N-octadecyl-3-aminopropyl-trimethoxysilyl chloride (DMOAP) for homeotropic surface boundary conditions. Dilute colloidal dispersions were obtained by dispersing particles in a LC host either via solvent exchange or mechanical mixing. Particles were dispersed in a nematic LC at a low density of about $10^5$ mm$^{-3}$, which allows observation of an isolated particle in the field of view with a 100× objective. Colloidal dispersions in a LC state were filled into LC cells after ~5 min sonication to break apart pre-existing aggregates. LC cells were assembled using glass substrates with transparent indium-tin-oxide (ITO) electrodes facing inwards to allow application of electric field between opposite substrates. To minimize spherical aberrations in experiments involving high numerical aperture (NA) immersion oil objectives, one of two substrates was 0.15-0.17 mm thick (SPI Supplies). Homeotropic surface boundary conditions at confining substrates were set by thin films of spin-coated and cross-linked polyimide SE1211 (Nissan Chemical Industries, Ltd.). To break the azimuthal degeneracy of the director's tilt during switching of homeotropically aligned nematic ZLI-2806 with negative dielectric anisotropy $\Delta\varepsilon=-4.8$, SE1211-coated substrates were 1-2 times rubbed with a velvet cloth applying a weak pressure of ~380 Pa along a direction $e_r$ (Figure 1a). Substrates with rubbed alignment layers were assembled in an antiparallel fashion so that $e_{r,t}$ and $e_{r,b}$ respectively on top and bottom substrates are parallel but pointing in opposite directions (Figure 1a). A cell gap thickness of $d≈5.5$ $\mu$m was set by glass spacers (Duke Scientific) dispersed in ultraviolet-curable glue (Norland Products, Inc.).

**Methods and Techniques.** Analog function generator GW Instek GFG-8216A or a data acquisition board (NIDAQ-6363, National Instruments) and a homemade MATLAB-based software were used for generating different driving schemes and waveforms of voltage applied to LC. The latter was used to obtain waveforms of low frequency voltage with a high carrier frequency $f_c$=1 kHz. High carrier frequency was used to avoid effects associated with a transport of ions at low-frequency applied electric fields. The homemade software allowed us to tune various parameters of applied voltage and generated waveforms were controlled using a Tektronix TDS 2002B oscilloscope. The threshold voltage $U_{th}$=1.84 $V_{p-p}$ of the Fréedericksz transition for a pure ZLI-2806 in homeotropic cells was experimentally determined from the dependence of a capacitance of the cell on the applied voltage measured at 1.0 kHz using an impedance gain-phase analyzer Schlumberger 1260.

An experimental setup assembled around an inverted Olympus IX81 microscope was used for optical bright-field and polarizing microscopy observations with 20× (NA=0.4) and oil 100× (NA=1.4) objectives. Dynamics of LC textures and translational motion of colloids was recorded using a charge-coupled device (CCD) camera (Flea, PointGrey) at a rate of 15 and 30 frames per second and the exact spatial positions of colloidal spheres as a function of time were then determined from captured video using motion tracking plugins of the ImageJ (freeware from NIH) analyzing software.

**RESULTS AND DISCUSSION**

**Colloid-hedgehog elastic dipole in a homeotropic nematic cell**

Elastic dipoles with a point defect called a hedgehog were obtained by mixing silica spherical particles (SPs) with homeotropic boundary conditions into a nematic LC and filling a resulting dispersion into a homeotropic LC cell (Figure 1a) of thickness $d \approx 5.5$ $\mu$m, which is slightly larger than a diameter $2R_0 = 3.14$ $\mu$m of colloidal particles. Dipolar configuration of the director $\mathbf{n}(\mathbf{r})$ around SPs[22] results from the mismatch of the preset far-field director $\mathbf{n}_0$ in the LC cell and the alignment of LC molecules at the particle's surface with homeotropic conditions. As can be seen from the schematic representation of $\mathbf{n}(\mathbf{r})$ in Figure 1a, SP is topologically equivalent to a radial hedgehog defect with a topological charge $s=+1$ and is accompanied by a hyperbolic hedgehog defect of opposite topological charge $s=-1$. The director field $\mathbf{n}(\mathbf{r})$ surrounding an elastic dipole at small distances is tilted away from $\mathbf{n}_0$ (Figure 1a). This tilt is represented by a two-dimensional polar vector $\mathbf{c}$ defined as a projection of the thickness-averaged $\mathbf{n}(\mathbf{r})$ onto the $x$-$y$ plane and sketched with the help of "nail" symbols in Figure 1b. The tilt of $\mathbf{n}(\mathbf{r})$ around SP results in a radial distribution of $\mathbf{c}$ (Figure 1b) and, under crossed polarizers, elastic dipoles in a homeotropic cell are seen as SP surrounded by four bright lobes separated by dark regions where $\mathbf{c}$ at the SP's perimeter matches the orientation of either polarizer or analyzer (Figure 1c). Using a full wave ($\lambda=530$ nm) phase retardation plate with a slow axis $\gamma$ placed after the sample at 45° between crossed polarizers, one can determine orientation of $\mathbf{c}$ in a plane of the cell based on the interference colors in experimental micrographs. Interference colors in a LC texture are respectively blue or yellow when nonzero **c**∥γ or **c**⊥γ. Magenta color is seen in the regions where a nonzero **c** is parallel either to a polarizer or analyzer or where **n(r)**=**n**$_0$ and **c**=0.

Typically a vertical position of colloidal inclusions in LC cells is determined by the balance between elastic repulsion from the walls of confining substrates and gravitational force $F_g$=(4/3)$\pi R_0^3 \Delta \rho$g≈0.26 pN, where $\Delta \rho$≈1650 kg m$^{-3}$ is the difference between LC and silica densities, and g=9.8 m s$^{-2}$ is the gravitational acceleration. In our case, due to the tight confinement, the vertical position of SPs is determined mainly by elastic repulsion from the walls as the force of this elastic repulsion can reach tens of piconewtons[12] and the modestly small contribution of $F_g$ can be neglected. Due to asymmetry of dipolar configuration of **n(r)**-distortions with respect to the particle's equator, the elastic repulsion potential between an elastic dipole and a wall with homeotropic anchoring is higher when a hedgehog is facing a wall by[27] $\delta E_{wall}$=3$\pi K \alpha \beta R_0^5 d^{-4}$, where $K$ is an average elastic constant of a nematic LC, $\alpha$ and $\beta$ are parameters related to dipole and quadrupole elastic moments,[22] respectively. Using $K$≈12.7 pN for ZLI-2806, and[22,28-30] $\alpha \approx$ 2 and $\beta \approx$ 0.5, we find the difference $\delta E_{wall}$≈5000$k_B T$ for our case, where $k_B$ is a Boltzmann constant and $T$ is temperature. Therefore, centers of mass of SPs do not reside perfectly in the middle of the cell but are slightly displaced by ≤0.5 μm towards one of the substrates; the particle's pole that is the closest to a hedgehog is pushed from a nearest wall further away than another pole (Figure 1a).

At no electric fields applied to a LC cell, SPs are freely roaming in a plane of a cell due to Brownian motion (Figure 1d); displacements are independent from the in-plane direction (Figure 1e). The significant vertical displacements of SPs are prevented by a tight confinement in a very narrow gap. Even two-dimensional displacements of SPs in the plane of the cell are hindered by such tight confinement and strong elastic coupling with both substrates. A two-dimensional diffusion coefficient $D=3.81\times10^{-4}$ $\mu$m$^2$ s$^{-1}$ for SPs determined from their Brownian motion[1,31] (Figure 1f) in our thin cells was ~3 times smaller than a diffusion coefficient $D=12.65\times10^{-4}$ $\mu$m$^2$ s$^{-1}$ of the same particles but in a 10 µm thick cell (Figure 1f) and almost 5 times smaller than a diffusion coefficient $D_{\text{calc}}=18.7\times10^{-4}$ $\mu$m$^2$ s$^{-1}$ in an isotropic phase calculated for a defect-free SP moving in an isotropic liquid with a viscosity $\eta=75$ mPa s of LC using a Stokes-Einstein relation[1] $D_{\text{calc}}=k_{\text{B}}T/6\pi\eta R_0$. This indicates a strong role played by defects and elastic coupling of **n(r)**-distortions fields around SPs to confining substrates, consistent with previous works of Stark,[32,33] Poulin[31] and others[34-39] and even further demonstrating the role of confinement.

**Electric field induced colloid-umbilic topological dipole**

In our experiments, we used nematic LC with negative dielectric anisotropy $\Delta\varepsilon=-4.8$ so that the director **n(r)** orients orthogonally to an applied electric field. When alternating current (AC) voltage $U$ (Figure 2a) larger than a threshold voltage $U_{\text{th}}=\pi(K_3/\varepsilon_0|\Delta\varepsilon|)^{1/2}$, where $K_3$ is a bend Frank elastic constant, is applied to ITO electrodes (Figure 2b), the director far from the particle rotates from its initial homeotropic orientation to tilted or even planar orientation depending on the voltage value because of LC's negative dielectric anisotropy (note that the deformation of director corona around the particle is threshold-free and takes place even at very small voltages). The macroscopically homogeneous direction of **n(r)** tilt and corresponding **c** is predetermined by a small (1-2°) pretilt with respect to homeotropic alignment induced by a weak rubbing[40] along $e_{r,t}\neq e_{r,b}$ (Figures 1a and 2b). The director **n(r)** tilts in the direction of rubbing $e_r$ (Figure 2b) as was confirmed by means of conoscopic observations.[41,42]

Figure 2e,g shows textures corresponding to the dynamics of **n(r)**-configurations around an elastic dipole upon applying an electric field. Let us consider, for example, the SP closer to the bottom substrate (top schematic of Figure 2b). An applied voltage $U>U_{th}$ causes homogeneous tilt of **n(r)** in the cell with respect to $U$ amplitude and rubbing $e_{r,b}$ with the in-plane **c** in the cell far away from SPs oriented along $e_{r,b}$. However, **n(r)** around the SP is already tilted away from **n**$_0$ even before applying voltage because of homeotropic boundary conditions at a particle's surface (Figure 1a), and this tilt is radially symmetric with respect to **n**$_0$ resulting in a radial **c** around SP (Figure 1b) with the direction of a polar **c** director in a plane $z$-$e_{r,b}$ being opposite on the left and right side of SP. When a moderate voltage is applied, this preexisting tilt changes its magnitude accordingly to the strength of an electric field, but it does not change its direction with respect to $z$-axis as there is no preference in azimuthal orientation of the tilt. Therefore, at one side of an elastic dipole along $e_{r,b}$, **n(r)** is tilted in the direction opposite to overall tilt in the cell (Figure 2b). This mismatch in tilt and corresponding **c** directions in the cell and around an elastic dipole results in appearance of a defect with four dark brushes in front of an elastic dipole with three of them closing on the surface of a particle (Figures 2d,e,g). These defects are umbilics,[2,43] regions where **c** rotates by $2\pi$ (Figure 2d), caused by the rotational degeneracy of the director tilt around an elastic dipole. The director field **n(r)** is continuous everywhere within an umbilic and there are no singularities. Umbilics have a continuous effective core with a radius[43] $r_c \sim (d/\pi)(U_{th}^2/[U^2-U_{th}^2])^{1/2}$. When $U<U_{th}$, the core of umbilics spreads out and the originally spatially localized umbilic becomes undefined (Figure 1a). The umbilic's continuously deformed core contracts into a small region where four dark brushes meet (Figure 2c-e) when $U>U_{th}$ and $r_c$ can be determined from the optical microscopy textures. The detailed **c**-field around the colloid-umbilic pair (Figure 2d) can be deduced from polarizing optical microscopy

textures using an additional retardation plate (Figure 2e,g). When $U>U_{th}$, there are two integer-strength self-compensating defects in the 2D in-plane **c**-field: one of strength $k=-1$ within an umbilic and another one is with a particle itself in its core topologically equivalent to a defect with strength $k=+1$ (Figure 2d). The topological defects in the **c**-field are well pronounced at $U>U_{th}$ because the radial distribution of **c** around the elastic dipole formed by SP and accompanying hedgehog has to be embedded into the homogeneous **c**-field caused by an applied $U$ and pre-determined by the rubbing direction.

The observed periodic transformations of **n(r)**-deformations around SPs due to the applied voltage oscillating between minimum and maximum values are also interesting from the perspective of classification of elastic multipoles in nematostatics.[25,26] At no voltage applied, the SP and corona of surrounding **n(r)**-deformations form an well-known isotropic or uniaxial elastic dipole[25,26] with $C_{\infty v}$ symmetry where **n(r)**-deformations are radially symmetric with respect to a line parallel to $\mathbf{n}_0$ going through the center of SP and a hedgehog (Figure 1a). When voltage is applied and an umbilic defect appears in the vicinity of SP, the uniaxial dipole continuously transforms into a general biaxial (nonchiral) dipole of $C_{1v}$ symmetry. This $C_{1v}$ elastic dipole has a vertical mirror plane $z$-$e_r$ describing the only remaining symmetry operation bringing the structure into itself, apart from the trivial operations, like the 360 degrees rotation (Figure 2b,c). This biaxial elastic dipole combines symmetry breaking along directions of the cell normal and in the plane of the cell along the rubbing directions. From the standpoint of view of topology, the structures of colloid-hedgehog dipole in **n(r)** and colloid-umbilic dipole in a 2D vector field structure of **c** are consistent with topological charge conservation laws associated with defect embedding in a uniform 3D and 2D field backgrounds, respectively, in a manner that complies with the constraints of topological theorems.[44] The configuration of the director field around the SP resembles that of a radial hedgehog and is a topologically nontrivial structure within the homotopy group $\pi_2(\mathbb{S}^2/\mathbb{Z}_2)=\mathbb{Z}$, compensated by a hyperbolic hedgehog belonging to the same homotopy class and having an elementary charge of an opposite sign (as can be seen upon vectorizing director[45]). In the 2D **c** director-tilt vector field, both the particle-induced and compensating field configurations belong to the first homotopy group, $\pi_1(\mathbb{S}^1)=\mathbb{Z}$, elements and self-compensate when embedded in the uniform far-field **c**. The reported here interesting field-induced transformations may find practical utility from the standpoint of new forms of controlled colloidal self-assembly.

**Electric field driven locomotion of colloids**

We used two waveforms of AC voltage in our experiments with the LC cells hosting elastic dipoles (Figure 2a). We applied either a sine-wave voltage at low frequency of $f_m$=1-5 Hz or a square-wave with high carrier frequency $f_c$=1 kHz additionally modulated by a sine wave of a low frequency $f_m$=1-5 Hz. The former waveform was also asymmetric with respect to a zero level due to our design that was based on combining it with a DC offset. In the latter waveform, having a high carrier frequency $f_c$ allows avoiding the effects associated with a transport of ions or ionic impurities at low-frequency applied electric fields and assuring only dielectric response of LC. The important feature of used waveforms is that voltage magnitudes in each half cycles are different so that $0<U_2/U_1<1/2$ (Figures 2a and 3c). When AC voltage of such waveforms is applied to the cells with elastic dipoles, they move in both lateral directions along $e_r$ (Figure 2c) with respective change of **n**(**r**)-configurations around them (Figure 2e,g) that depends on $f_m$ (Figure 2f). As already mentioned above, due to asymmetry of the 3D dipolar **n**(**r**)-configuration, elastic dipoles are residing slightly closer to one of the substrates (Figure 1a). Thus, the vertical position of elastic dipoles in the cell and the particle-hedgehog dipole orientation of the elastic dipole moment determine not only the positions of induced umbilics but also the direction of particles' translational motion. SPs effectively move in the direction of $e_r$ at the nearest wall so that SP near the bottom substrate moves along $e_{r,b}$ and SP near the top substrate moves along $e_{r,t}$ (Figure 2b,c). The directional motion of elastic dipoles is stepwise (Figure 3a,b). The particle advances every half cycle of the applied periodic voltage signal, when $U=U_1>U_{th}$, and rests nearly motionless for another half cycle when $U=U_2$ (Figures 2a and 3a,c). This jerky motion results in a net directional locomotion and displacement $L$ of elastic dipoles with a net velocity $v_{net}>0$ (Figure 3a). The asymmetry of an applied waveform with different voltage magnitudes $U_2 \neq U_1$ in each half period is important for the efficient motion of elastic dipoles. Indeed, when a symmetrical sine wave voltage is applied, particles' net displacement is equal zero (Figure 3a).

During the first half cycle of an oscillating signal an amplitude of applied voltage is higher than the threshold value $U_{th}$, an umbilic appears in front of the SP at some distance $d_{pd}$ (Figure 2d) separated along $e_{r,b}$. At first, the umbilic defect appears at the maximum distance (for a given modulation period) $d_{pd,max}$ from the SP (Figure 2d), which depends on the magnitude of $\mathbf{n}(\mathbf{r})$-tilt caused by $U$. As the voltage signal evolves with time, the umbilic defect and colloidal SP particles move swiftly toward each other over a short period of time when $U>U_{th}$. Interestingly, the motion of SP does not continue all the time while $U>U_{th}$ but only within a short period of time right before $U$ reaches the maximum (Figure 3c), resulting in a jerky motion of a particle. The ensuing stepwise jerky displacement of the particle is directed along the particle-umbilic separation vector and repeats every other half period resulting, as time progresses, in a directional motion of the particle with some $v_{net}$ (Figure 3b-d). Qualitatively, the temporal evolution of colloidal and director structures associated with the motion of an elastic dipole with accompanying dynamics of $\mathbf{n}(\mathbf{r})$-structure (Figure 3c and Video S1) resembles a swimmer moving with help of a butterfly stroke, albeit here this colloidal motion takes place at low Reynolds and Ericksen numbers and is possible primarily because of the nonreciprocity of the director field evolution. Indeed, the temporal out-of-equilibrium evolutions of the director and **c**-field distortions around particles upon turning voltage larger and smaller than $U_{th}$ are not invariant upon reversal of time, as can be seen from the polarizing optical micrograph sequences (Figure 2e,g), prompting lateral translations of SP in the nematic medium of a low Reynold's number. The particle's net period-averaged velocity can reach $v_{net} \approx 1$ µm s$^{-1}$ depending on applied voltage parameters but the instantaneous velocity $v_s$ during each step or stroke is several times larger (Figure 3d); for example, for $U_1$=4 V$_{p-p}$ at $f_m$=1.8 Hz, the stroke velocity was measured to be $v_s \approx 3.6$ µm s$^{-1}$. Key to the observed phenomena is the dynamics of the umbilical structure, which tends to localize at a well-defined distance from the SP within each modulation period, and then interact with the SP at high instantaneous voltage magnitude but spread smoothly at no applied field. At high instantaneous applied fields, the elastic dipole and umbilic are driven towards each other by elastic attraction forces between corresponding topological defects of equal integer strength and opposite sign in the **c**-field (Figure 2d), which minimizes their contribution to the free elastic energy. Experimentally measuring $v_s$ and $D$=3.81×10$^{-4}$ µm$^2$ s$^{-1}$ for an elastic dipole (Figure 1b,d), one can determine a force $F_{el}$ pulling on SP during elastic attraction between the umbilic and elastic dipole from balancing it with a viscous drag force $F_S$=($k_BT/D$) $v_s$ as a Reynolds number is $Re=\rho v_s R_0/\eta \approx 10^{-8}$<<1 at the LC density $\rho \approx 10^3$ kg m$^{-3}$. The measured $F_{el}$=20-50 pN is quite large and comparable to the magnitudes of forces of elastic attraction between elastic multipoles in nematics.[46-50] As a result of this elastic attraction, SPs are propelled by applied voltage in a well-defined direction along $e_r$ (Figure 2c). Within the fractions of each modulation period when instantaneous voltage is low or zero, the umbilic "spreads" to become a delocalized structure of weak distortions that does not strongly interact with the SP. Thus, the nonreciprocity of director field evolution with modulation can be also linked with and analyzed in terms of the asymmetry of elastic forces acting on the SP. From this direct experimental probing of elastic interactions between an elastic dipole and the induced umbilical structure, it is clear that the observed voltage-controlled directional motion of SP does not stem from hydrodynamic effects, electroconvection, backflows or related phenomena involving the actual LC flows induced by fields, albeit it certainly can be somewhat influenced by these effects and flows are certainly generated by the particle as a consequence of its motion. Backflows associated with electric switching of director cause negligible effects on the particle displacements as the elastic dipoles are held nearly in the middle of the cell due to the very tight confinement in a narrow cell gap almost comparable with the particle size. Unlike in the previous works on electrically driven and backflow-assisted particle motions,[15] our SPs are not substantially displaced from the cell middle towards substrates, where the nematic fluid displacement due to the backflow is maximized upon application of an electric field. Furthermore, the Ericksen number $Er=\eta v_s R_0/K \approx 0.004\text{-}0.08$ at a LC viscosity $\eta \approx 1\text{-}300$ mPa s (covering typical values of LC's viscosity coefficients) is much smaller than unity, indicating that the flows do not couple strongly with the elastic deformations, as well as the dominance of elastic forces over the hydrodynamic forces[32,33,39] during the SP's locomotion.

To further elucidate any possible critical role that could be played by ionic currents and associated hydrodynamic effects, we used waveforms of applied voltage with a high carrier frequency $f_c$=1 kHz (Figures 2a,f and 3c). The fact that the motion is found unimpeded when using this high carrier frequency shows that ionic effects are not critical for the observed phenomena. The high carrier frequency corresponds to electric field direction oscillations/inversions within characteristic times $1/f_c$ short enough to preclude significant ion motions, supporting further the physical origins of the motion described above. Net velocity of elastic dipoles can be varied by varying a modulation frequency $f_m$=1-5 Hz, a magnitude of applied voltage $U$ and $0.25 \leq U_2/U_1 < 0.5$ (Figure 4). Dependence of $v_{net}$ on $f_m$ or $U$ is non-monotonous (Figure 4b,f): $v_{net}$ increases quadratically with increasing $U$ at a constant $f_m$ but then quickly decreases after applied voltage reaches modestly high values (>10 V$_{p-p}$) at which the director in the middle plane of the cell is reoriented almost parallel to the substrates and an umbilic defect does not appear anymore in front of SP. At even higher applied voltages, the effectively immobile elastic dipole with typical **n(r)**-deformations[48-50] is observed in the nematic bulk with in-plane orientation of the far-field director across most of the cell thickness. Interestingly, there is a nonlinear correlation between $v_{net}$ and a maximum distance between an elastic dipole and umbilic $d_{pd,max}$ upon its appearance, which is decreasing with increasing either $f_m$ or $U$ (Figure 4a,c,d,e). It appears to be related to an optimal distance for elastic interactions between an elastic dipole and umbilic and their duration determined by a frequency of applied voltage.

While our experiments provide insights into the importance of the nonreciprocal rotational dynamics of director field in powering colloidal locomotion, the exact contributions and interplay of various effects in defining electrically driven nematic colloidal motions under different conditions in both our experiments presented here and in the past studies[13-19] will need to be explored further. For example, recently Long and Selinger have theoretically confirmed[51] that electric field induced motion of particle-like topological solitons can be explained by the nonreciprocal evolution of the director field, consistently with the earlier experimental and numerical findings.[20] Extensions of such theoretical models from purely solitonic field configurations to the more complex case of nematic colloids could provide particularly valuable insights into the relative contributions of nonreciprocal director dynamics, elastic interactions, backflow, electrokinetic and other effects into defining and controlling nematic colloidal locomotion under various experimental conditions.[13-19]

**CONCLUSIONS**

We have studied colloidal particles with homeotropic boundary conditions in thin homeotropic nematic cells and demonstrated electric-field-assisted formation of a dual nature elastic/topological dipoles consisting of an always-present, conventional dipole of colloid-hedgehog pair and a dynamically induced 2D topological dipole formed by a colloidal particle and an umbilic defect in the tilt directionality field. The nonreciprocal dynamics of a director field transformation allows for an electrically controlled locomotion of colloidal particles and can be analyzed in terms of LC-mediated interactions between the colloidal particle and an umbilic defect. We have shown that the dynamics of the director field transformations depends both on an amplitude and frequency of applied voltage, with nonlinear dependencies of locomotion velocity of colloids on these parameters. Our results indicate that the nonreciprocal "squirming"-like evolution of particle-induced director structures and defects plays an important role in defining electrically powered dynamics and locomotion of colloidal particles in an oscillating field, similar to that also reported for topological solitons.[20,51] Our findings can be used to develop different driven, active and other out-of-equilibrium self-reconfigurable systems, microfluidic transport systems, microrobotics[52] and various technological applications. From the fundamental perspective, the described rich dynamic phenomena provide new insights into nematic colloids, their assembly, nonequilibrium reconfiguration behavior, and collective dynamics.

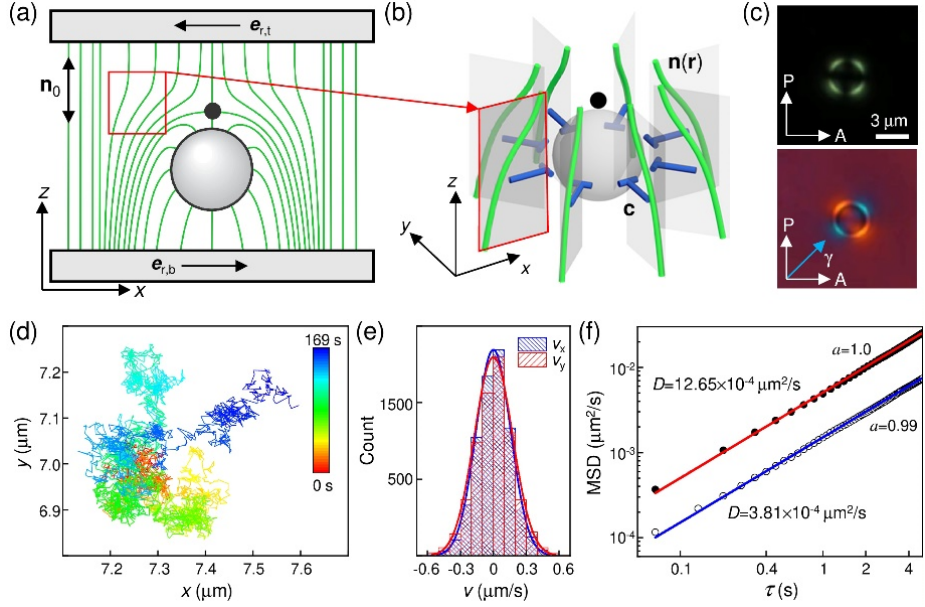

**Figure 1.** Elastic dipoles in a homeotropic nematic cell: (a) Schematic of a director field (green lines) around an elastic dipole in a homeotropic cell; $e_{r,t}$ and $e_{r,b}$ show a direction of weak rubbing of respectively top and bottom confining substrates coated with alignment films. Note that the schematic is not to scale as the particle diameter in our experiments constitutes ~60 % of cell thickness. (b) Definition of a polar vector **c** (blue nails), which is a projection of a tilted director **n(r)** on *x-y* plane. The LC-embedded SP with perpendicular anchoring induces radial 2D tilt directionality **c** vector field with a radial orientation close to the particle. (c) Polarizing microscopy textures of an elastic dipole in a homeotropic cell. (d) Example of a trajectory of SP undergoing Brownian motion in the plane of a homeotropic cell. (e) Histograms of velocities during displacements in the plane of a cell along *x* and *y* directions. (f) Log-log plots of mean-square-displacement (MSD) as a function of lag time measured for SP in a homeotropic cells of different thickness *d*=5.5 mm (blue fitting line) and *d*=10 μm (red fitting line). Experimental data were fitted using an expression[1] MSD=$4D\tau^a$ with $a \approx 1$, which indicates normal diffusion of SPs due to random Brownian motion.

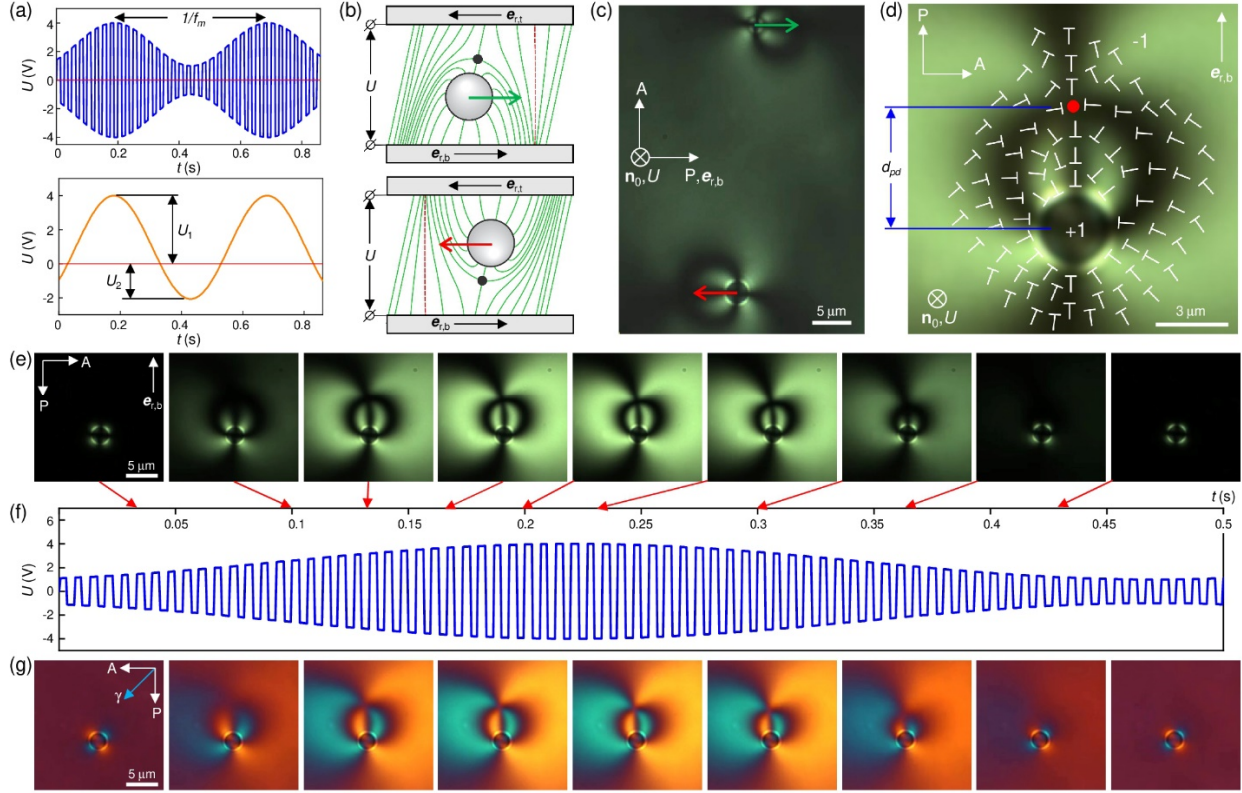

**Figure 2.** Locomotion of elastic dipoles in a homeotropic nematic cell: (a) Waveforms of an applied voltage $U$. (b) Schematic of a director field around two elastic dipoles moving in different directions in (c). Note that the schematic is not to scale as the particle diameter in our experiments constitutes ~60 % of cell thickness. (c) Polarizing microscope texture around two elastic dipoles moving under an applied voltage $U$ in different directions. (d) Schematic of a **c**-field around an elastic dipole and umbilic defect under the applied voltage. (e-g) Sequences of polarizing microscope textures around an elastic dipole obtained without (e) and with (g) a retardation plate changing with respect to an applied voltage with $f_m$=2 Hz and $f_c$=1 kHz (f); for demonstration purpose a carrier signal in (f) is represented by lower frequency.

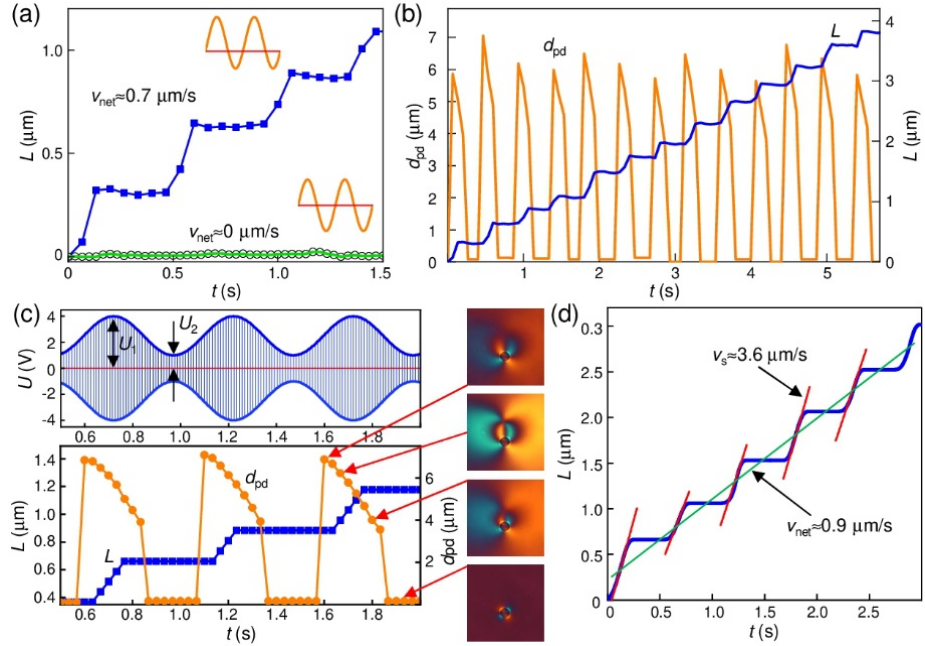

**Figure 3.** (a) Displacement of an elastic dipole over time at asymmetric and symmetric waveforms of applied voltage. (b) Distance between an elastic dipole and umbilic defect and corresponding displacement of an elastic dipole at $f_m$=2.25 Hz. (c) Applied voltage waveform with $f_m$=2 Hz and $f_c$=1 kHz and corresponding distance between an elastic dipole and umbilic defect (blue filled circles) and displacement of an elastic dipole versus time (red filled squares). Insets show corresponding **n**(**r**)-textures obtained between crossed polarizers and a full wave retardation plate. (d) Plot of an elastic dipole's locomotion versus time (a blue line) showing details of stroke and net displacements. Red and green lines are linear fits for stroke and net displacements, respectively.

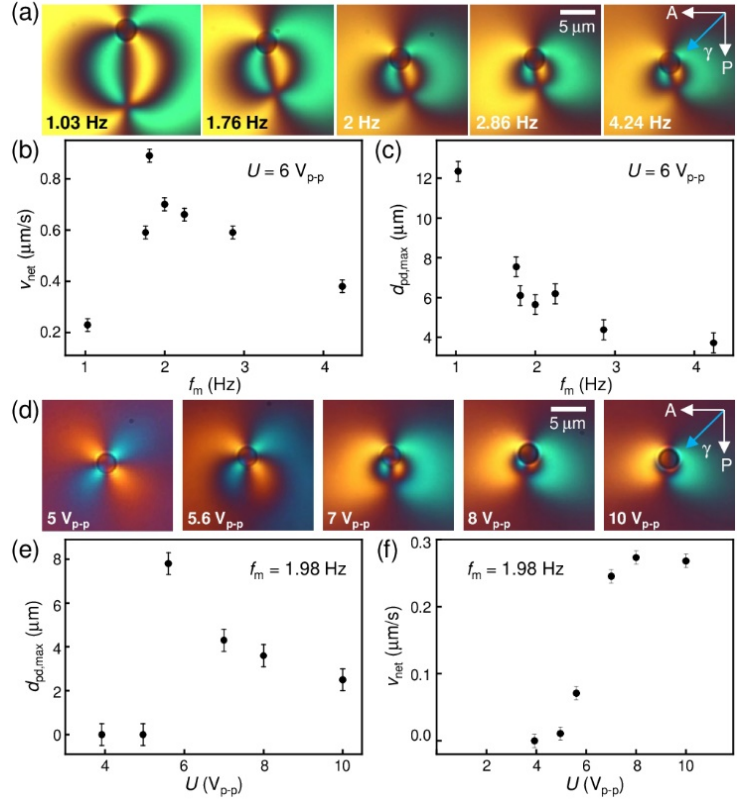

**Figure 4.** (a) Textures of **n**(**r**) obtained between crossed polarizers and a full wave retardation plate showing the distance $d_{\text{pd,max}}$ between an elastic dipole and umbilic defect depending on the frequency $f_{\text{m}}$ of applied voltage. (b) Velocity of SP and (c) distance $d_{\text{pd,max}}$ depending on the frequency $f_{\text{m}}$ at a constant applied voltage $U=6$ V$_{\text{p-p}}$. (d) Textures of **n**(**r**) showing $d_{\text{pd,max}}$ between an elastic dipole and umbilic defect depending on applied voltage at constant frequency $f_{\text{m}} \approx 2$ Hz. (e) Distance $d_{\text{pd,max}}$ and (f) velocity $v_{\text{net}}$ of SP depending on applied voltage at constant frequency $f_{\text{m}} \approx 2$ Hz.

**Author Contributions**

B.S. and R.E.A. conducted the experiments. I.I.S. conceived and directed the project. B.S. and I.I.S. analyzed the results. The manuscript was written through contributions of all authors. All authors have given approval to the final version of the manuscript.

**ACKNOWLEDGMENT**

We acknowledge discussions with H. Mundoor, J. B. ten Hove, A. Repula. This research was supported by the National Science Foundation through grants DMR-1810513.